# IMPACT OF MOBILITY ON POWER CONSUMPTION IN RPL

By

CHANDRA SEKHAR SANABOINA *    PALLAMSETTY SANABOINA **

\* Research Scholar, Jawaharlal Nehru Technological University, Kakinada, Andhra Pradesh, India.
\*\* Professor, Department of Computer Science and Systems Engineering, Andhra University, Visakhapatnam, Andhra Pradesh, India.

*ABSTRACT*

*The main theme of this paper is to implement the mobility model in Cooja simulator and to investigate the impact of the mobility on the performance of Routing Protocol over Low power Lossy networks (RPL) in the IoT environment. In the real world, mobility occurs frequently. Therefore in this paper, a frequently used mobility model - Random Way Point (RWP) is used for analysis. RWP can be readily applied to many existing applications. By default, the Cooja simulator does not support mobility models. For this, the Bonn Motion is introduced into Cooja as a plugin. As IoT deals with the resource-constrained environment, a comparison is done between the static environment and the mobile environment in terms of power consumption. As expected, the results indicate that mobility affects the RPL in terms of Power Consumption.*

*Keywords: IoT, RPL, Mobility Models, Power, Resource Constrained Environment, Cooja Simulator, RWP.*

## INTRODUCTION

Till now the internet is being used for browsing the web, accessing multimedia content, playing games, social networking, and topic search, sending and receiving emails and in many other tasks and now the trend is shifting towards usage of the internet as a global platform for communication between machines (M2M). Within the next few years, the Internet will turn as a seamless platform for traditional networks as well as networked objects thus paving a way to a new era of Interconnected Smart Objects forming Pervasive Computing Environments (Weiser, 1999).

This does not mean that Internet infrastructure will disappear. It will continue its role as a global backbone for WWW but in addition, extends its hand towards interconnecting physical objects with computing and communication capabilities across a wide range of services and technologies (Smart Objects). This can be achieved by embedding electronics into physical objects making them Smart Objects. Smart objects (or things) can be defined as the entities (Miorandi et al., 2012) that have a physical embodiment and a set of associated physical features (e.g., size, shape, etc.), and have a minimal set of communication functionalities, such as ability to be discovered and to accept incoming messages and reply to them, possess a unique identifier, associated to at least one name and one address, possess some basic computing capabilities, may possess some means to sense physical phenomena (e.g., temperature, light, electromagnetic radiation). The term Internet of Things was first coined by Kevin Ashton in 1999 in the context of supply chain management (Ashton, 2009). The three pillars of IoT are:

- Being identifiable,
- Is being communicable, and
- Being interactive (ability to interact with anything).

The Internet-of-Things can be treated as a highly dynamic distributed networked system with a large number of smart objects that are capable of producing and consuming highly dynamic information. The vision of IoT provides opportunities to manufacturers and companies including users. It will find wide applicability in many sectors, such as agriculture, environmental monitoring, health-care, product management, inventory management, home automation, transportation, and logistics domain, supply chain management, disaster alerting and recovery, utilities, enterprise, security, and surveillance.

The features of the IoT can be identified as Device heterogeneity, Scalability, Ubiquitous data exchange, Power-optimized solutions, Localization and tracking capabilities, Self-Organization Capabilities, Semantic Interoperability, Privacy-Preserving, and Security. Internet of Things (IoT) ecosystem is a complex environment consisting of much heterogeneous hardware as well as software components. A large amount of data will be generated by sensors in the real world and hence impose a great demand for data storage and data processing which can later be converted into useful services or information. Some applications need very complex processing procedures that include historical data and time series analysis, whereas some applications are sensitive to latency and some applications are very simple in nature. Therefore, it is very difficult to imagine a real-world and ultra-scale IoT system without including a Cloud or some powerful devices.

In IoT, a resource may go from physical resources, for example, memory, CPU, Network Bandwidth, Power and so forth, to programming resources like virtualization function, strategies that perform data combination, or methods to distinguish a mind-boggling occasion, and so on. In cloud computing systems, the provision of service is pay per use, dynamic and elastic model. This depends on the formal or semiformal contracts between the customer and the cloud supplier. The resource allocation involves complicated algorithms to allocate better physical or virtual resources to applications.

In IoT, the necessities of resource distribution mechanism for cloud computing holds great yet with some extra prerequisites like adapting up to the ad-hoc nature of IoT and astute ad-hoc cooperations among gadgets and clients. Interactions for the administration arrangement in IoT makes the procedure of resource allotting and administration more intricate than in customary cloud computing. IoT frameworks handle hundreds, thousands, or a large number of parallel solicitations and a few kinds of utilization request quick reaction, inside a strict time interim. In IoT, multiple applications with potentially different requirements will be sharing the same resources (Delicato et al., 2017). Priority in access to shared resources should be given to time-critical applications, whereas in non-critical applications should guarantee that its requirements are met. Additionally, the nature of information created by IoT gadgets likewise influences its handling. Sensors deliver a tremendous amount of data which can be data stream, which ranges from a couple of bytes for every second to a couple of gigabits for every second. This information rate can be sporadic, unusual, and bursty in nature. Dealing with the resources associated with handling IoT data and conveying IoT administrations need to take into account the following issues: resource constraints like Computational constraints, Storage problems, Bandwidth Constraints, Application-aware protocols/Context-aware protocols, and Infrastructure support.

## 1. Literature Review

### 1.1 RPL

RPL (Winter et al., 2012) is a routing protocol for low power and Lossy networks designed by the IETF Routing over Low power and Lossy network (ROLL) group, utilized as the current routing protocol in Contiki. RPL gives a mechanism to disperse data over the powerfully shaped network topology. This dissemination empowers negligible design in the nodes, enabling nodes to work generally self-ruling. RPL fundamentally underpins multipoint-to-point traffic, with sensible help for point-to-multipoint traffic and essential highlights for point-to-point traffic. It operates under the assumption that the network contains a sink node with greater computing ability and power resources than the rest of the nodes in the network.

The initial phase in the network registration is Neighbour Discovery (ND). This causes the node to decide the neighbors in the region and to choose the best parent accessible. The node will initially transmit an RS (Router Solicitation) packet as a multicast to each and every routers. On getting the RS packet, every one of the routers reacts back with an RA (Router Advertisement) as a unicast to the node. The RA packet contains Prefix Information (PIO), Context Option (CO), and Authoritative

Border Router Option (ABRO).

After accepting the RA, the node chooses a router as its default router (in light of first got RA) and infers the worldwide IPv6 address in view of the prefix option. The node at that point sends a Neighbour Solicitation (NS) as a unicast message to its default router. The NS will contain the Address Registration Option (ARO). This option will tell the switch that the hub is straightforwardly reachable and furthermore the connection layer address of the node. The switch will influence a passage of the node in its Neighbour to reserve and react with a Neighbour Advertisement (NA) with the status of address enrolment.

After Neighbour Discovery, RPL is introduced and the network enrolment process will start (Tsvetkov & Klein, 2011). RPL builds a graph known as Destination Oriented Directed Acyclic Graph (DODAG). The whole network topology will be divided into multiple RPL instances. These RPL instances is an arrangement of different DODAG. Each DODAG is extraordinarily recognized by DODAGID (Winter et al., 2012).

- *Directed Acyclic Graph (DAG):* A directed graph having the property that all edges are arranged such that no cycles exist. All edges are contained in paths situated toward and ending at least one root nodes.

- *DAG root:* A DAG root is a node inside the DAG that has no outgoing edges. Since the graph is non-cyclic, by definition, all DAGs must have no less than one DAG root and all paths end at a DAG root.

- *Destination-Oriented DAG (DODAG):* A DAG established at a solitary destination, i.e., at a solitary DAG root (the DODAG root) with no outgoing edges.

- *DODAG root:* A DODAG root is the DAG base of a DODAG. The DODAG root may go about as a border router for the DODAG; specifically, it might total routes in the DODAG and may redistribute DODAG routes into other router protocols.

- *Virtual DODAG root:* A Virtual DODAG root is the after-effect of at least two RPL routers, for example, 6LoWPAN Border Routers (6LBRs), organizing to synchronize DODAG state and act in the show as though they are a solitary DODAG root (with different interfaces), as for the LLN. The co-ordination doubtlessly happens between controlled gadgets over a solid travel interface.

- *Up:* Up alludes to the heading from leaf nodes towards DODAG roots, following DODAG edges. This is utilized in graphs and depth first-search, where vertices from the root are deeper or "down" and vertices closer to the root are "shallower" or "up".

- *Down:* Down alludes to the course from DODAG roots towards leaf nodes, reverse of DODAG edges. This takes is utilized as a part of graphs and depth first-search, where vertices advance from the root are deeper or "down" and vertices closer to the root are "shallower" or "up".

- *Rank:* A node's Rank characterizes the node's individual position in respect to different nodes concerning a DODAG root. Rank entirely increments in the Downward direction and entirely reduces in the Up direction. This upon the DAG's Objective Function (OF). The Rank may similarly track a basic topological distance, might be figured as an element of connection measurements, and may consider different properties, for example, constraints.

- *Objective Function (OF):* An OF characterizes how routing metrics, optimization objectives, and related functions are utilized to process Rank. Moreover, the OF directs how parents in the DODAG are chosen and, subsequently, the DODAG arrangement.

- *Objective Code Point (OCP):* An OCP is an identifier that demonstrates which Objective Function, the DODAG uses.

- *RPLInstanceID:* An RPLInstanceID is one of a kind identifier inside a network. DODAGs with the same RPLInstanceID share a same Objective Function.

RPL is the only one of the protocols presented, which may also employ source routing. This occurs when it is operating in non-storing mode (Herberg & Clausen, 2011), which gives a basic assessment of the RPL protocol. Among others, it records its firmness as far as data traffic, particularly point-to-point activity, conceivable control packet discontinuity and the suspicion of bidirectional connections as hazardous purposes of the specification.

*1.2 Power-Aware Metrics*

Metrics are utilized to evaluate the nature of a connection or route under specific angles. The most commonly sent metric is Hop Count, with which the route utilizing the least hops is picked. Be that as it may, this is frequently not as much as perfect: not all connections are made equivalent in quality, and long-distance links are particularly inclined to be lossy. Power-awareness might be acquainted with existing routing protocols with the assistance of appropriate metrics.

A metric which takes power levels on either the node or network level may impact the routing choices of a protocol in a way which preserves power resources. Vasseur et al. (2011) specify the several routing metrics for the Routing Protocol for Low Power and Lossy Networks (RPL) protocol, some of which may be interesting for other deployments as well, and the notable are:

- *Node Power:* The power level of a node might be considered in various ways: Most instinctively, it might be received to pick a route over nodes with extraordinary leftover power, keeping in mind the end goal to stretch its lifetime and alleviate nodes with fewer resources. In doing as such, the estimation of remaining power must be put into the setting by the transceiver expenses of the individual node and in addition its normal lifetime. It might be useful to utilize a node with less battery which is probably going to be energized sooner rather than later (e.g. a cell phone on the nightstand) than one with high remaining power that needs to keep going for some time (e.g. a node in the wall).

- *Throughput:* At the point when the information sent over a router surpasses the measure of throughput, it can deal with, the subsequent packet errors will cause retransmissions, squandering power on excess communication. Hence, a router may determine the throughput it can deal with.

- *Latency:* Distinctive kinds of data may have diverse latency constraints, for example, in light of the fact that the information may wind up rapidly, is critical if there should arise an occurrence of a crisis or may trigger timeouts. By considering these necessities, a convention can disseminate the network stack in a way that backings diverse traffic prerequisites. These methodologies can be joined: (Ortiz et al., 2013) have proposed the utilization of Fuzzy Logic to consolidate a few applicable attributes of a course or connection into one explanation about its quality.

2. Background

*2.1 6LoWPAN and Routing Protocol for Low Power Lossy Networks*

Internet Engineering Task Force (IETF) revealed the IPv6 over Low-Power Wireless Personal Area Networks (6LoWPAN). 6LoWPAN characterizes an arrangement of protocols that can be utilized to coordinate the sensor nodes with IPv6 systems. Some commercial products have already released the protocol suite with core protocols composing the 6LoWPAN architecture. An applicable IETF Working Group named Routing Over Low power and Lossy systems (ROLL) has recently created the RPL routing protocol draft, which is the reason for routing over low power and lossy systems including 6LoWPAN. The RPL protocol develops as the accepted IPv6 routing standard for WSN. It is a tree-arranged routing protocol to shape a Destination-Oriented Directed Acyclic Graph (DODAG) with some characterized measurements and an objective function to control the choice of the best way to the root node.

RPL provides a mechanism to disseminate information over the dynamically formed network topology by a set of ICMPv6 control messages, such as DIO, DAO, and DIS (Zhang & Li, 2014).

- DIO message contains information about the rank, the objective function, the node id, and so on. It defines and maintains upward routes.

- DAO message promotes prefix reachability towards the client nodes of a DODAG to empower downward traffic.

- DIS message is utilized to proactively request the DODAG related data from neighboring nodes.

Generated Data from the nodes will be received by the root node. The root node in RPL is like a connection between the wireless sensor network and the internet. RPL still needs a lot of contributions to reach a full solution. Some of which are power consumption without losing accuracy. As the sensor nodes have limited supply of power, Power consumption is a big issue in IoT. Hence, routing protocols for IoT should be designed such that it maximizes the power-consumption. Sensor nodes can act as a sender, receiver, or a router. Any malfunctioning of the sensor nodes is not accepted and hence the accuracy of the routing protocols is maintained in the presence of low power sensor nodes in another aspect of power concern.

### 2.2 Mobility Models

The mobility models are classified into two principles considering specific mobility characteristics of each model. In real life environment, the common mobility model is Trace Model which is the first mobility model to be considered. Traces give exact data, particularly when they include countless participants and a properly long perception period. The second mobility model is the synthetic model (Sanchez & Manzoni, 1999), which attempts to sensibly represent to the practices of MNs without utilization of traces. This model contains two types, viz., a group mobility model and an entity mobility model as shown in Figure 1.

A mobility model should attempt to copy the developments of real Mobile Nodes (MNs). Alters in speed and course should be insensible time slots. For instance, MNs need not travel in straight lines at consistent speeds over the span of the whole simulation, since real MNs would not go in such a limited way. In this paper, different synthetic entity mobility models for ad hoc networks are discussed.

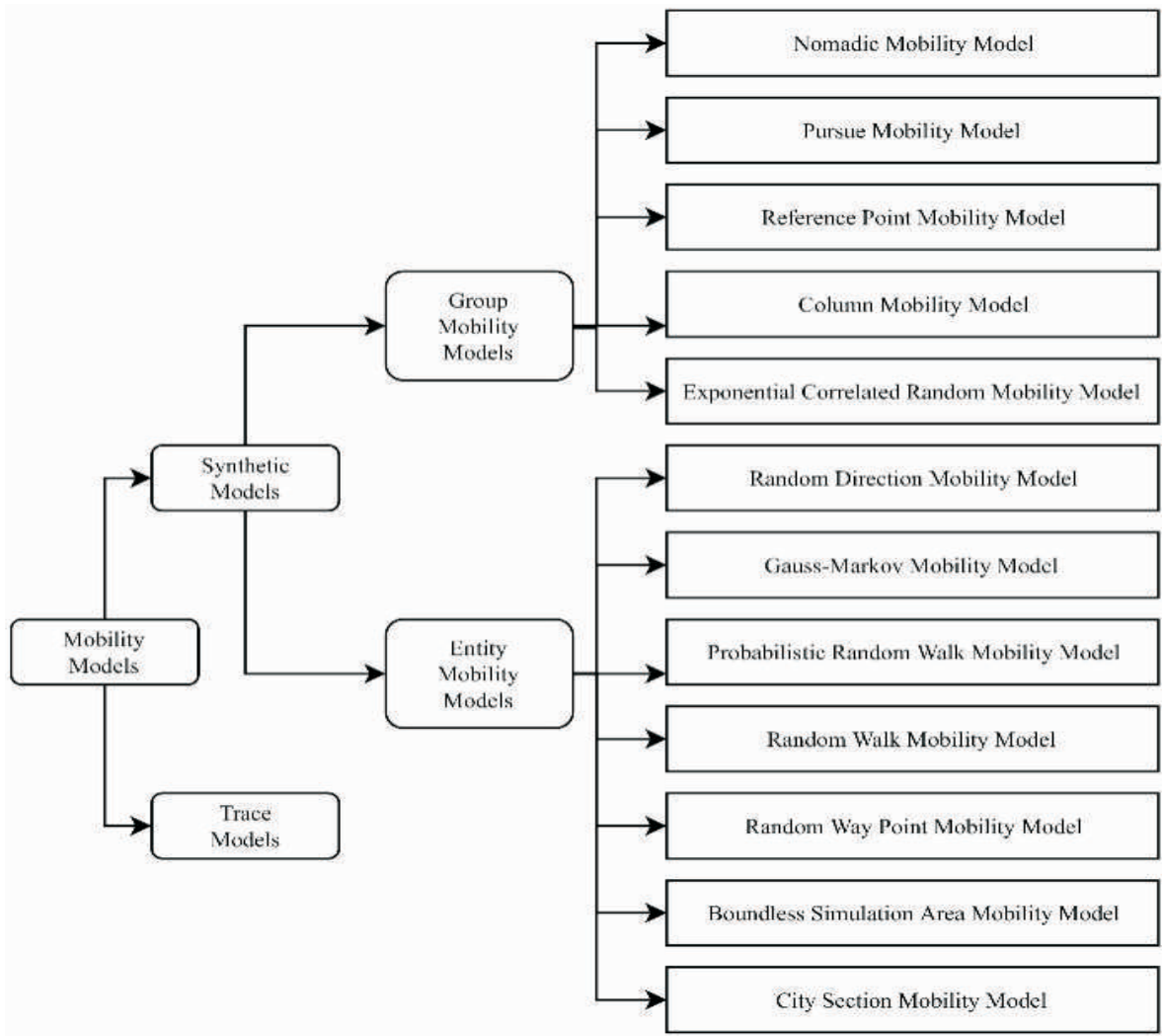

Figure 1. Structure of Mobility Models

*2.2.1 Entity Mobility Models*

Seven mobility models were proposed for (or utilized as a part of) the execution assessment of an ad-hoc network protocol. The principal model, the Random Way Point (RWP) Mobility Model is the most widely recognized mobility model utilized by specialists and RWP mobility model is utilized for investigation in this paper. Therefore, RWP is discussed in more depth than the other six models presented.

*2.2.2 Random Way Point (RWP) Mobility Model*

RWP mobility model is the famous model utilized by the exploration community as a result of its wide accessibility and its effortlessness to utilize. The RWP model was proposed by (Johansson et al., 1999; Broch et al., 1998). In simulation fields, the mobile node selects a position (x, y) randomly as a destination and choose randomly and uniformly the velocity from a range ($V_{min}$, $V_{max}$) to travel towards the destination.

When it reaches the destination, the node stops for little time called 'pause time' parameter, $T_{pause}$. After this timeframe, the node picks one progressively another destination randomly towards it, and a constant similar procedure is followed until the point when simulation closes.

In the RWP model, the mobility behavior of nodes will be decided by two main parameters: $V_{max}$ and $T_{pause}$. The topology of Ad-Hoc network will be in the stable mode when the $V_{max}$ is minimum and $T_{pause}$ is long. In a similar way, when $V_{max}$ is maximum and $T_{pause}$ is low, the ad-hoc network will be in dynamic mode (Bai et al., 2003). These two parameters play a major role in the RWP model. By varying these parameters, mainly the $V_{max}$ parameter, various mobility situations can be generated with various node speed. Subsequently, it appears to be imperative to measure the node speed.

One of the ideas is the average node speed. When $T_{pause}$ is 0, assuming $V_{max}$ is consistently and randomly chosen between [0, $V_{max}$], then the average node speed will be 0.5 $V_{max}$ (Camp et al., 2002). Be that as it may, the $T_{pause}$ parameter should not be ignored because it is the relative speed of two nodes that choose if the bridge between them breaks or forms, rather than their individual rates

(Kaur, 2012). In this way, average node speed gives off an impression of being the fitting metric to represent node speed. For instance, the development hint of a node is shown in Figure 2.

From (Johansson et al., 1999), the mobility metric to catch and measure this node speed with the relative speed between nodes i and j at time t is:

$$RS(i, j, t) = \left| \vec{V_i}(t) - \vec{V_j}(t) \right| \quad (1)$$

By then, the Mobility metric M is figured as the proportion of relative speed landed at the midpoint of overall node sets and in overall time (Kaur, 2012).

$$\bar{M} = \frac{1}{|i,j|} \sum_{i=1}^{N} \sum_{j=i+1}^{N} \frac{1}{T} \int_{0}^{T} RS(i, j, t) dt \quad (2)$$

where $|i, j|$ is the quantity of particular node match (i, j), n is the aggregate number of nodes in the simulation field, and T is the simulation time.

*2.2.3 Random Walk (RW) Mobility Model*

This mobility model was produced and portrayed numerically by Einstein in 1926 to copy the unpredictable developments of the particles known as Brownian motion. In this model, a node begins its movement by choosing a direction with speed from the pre-determined extents [0, 2*π] and [0, $V_{max}$]. The node moves for a settled time interval t or moves for a settled distance d. After distance d or time t, new direction and speed are chosen from the pre-determined extents (Camp et al., 2002).

RW mobility model acts comparatively to the RWP Mobility Model. In the two models, development of the node has solid randomness. Moreover, Nodes move along an unpredictable path and for this RW is recommended to mimic their development. Besides, nodes do not preserve their speed and direction, however it changes these two parameters in each time interval. It is considered as a memoryless mobility process because each step is calculated independently with the previous one.

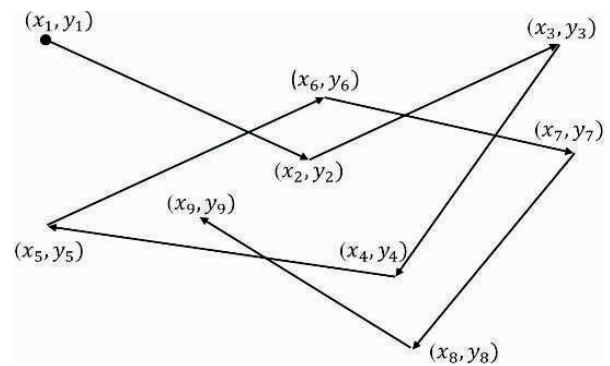

Figure 2. Example of Node Movement in the Random Way Point Model

*2.2.4 Random Direction (RDM) Mobility Model*

In this model, the node chooses direction and velocity from the range (0, 2*π) and (0, $V_{max}$). When node reaches at the end of the simulation area, it waits there for pause time. After the expiry of pause time, another direction will be chosen from (0, π) and continues to travel towards the end of the simulation area in a new direction. This process of choosing random direction and velocity continues until the completion of the simulation. RDM Model is similar to the RWM Model, but with a small difference like the motion of the node to the end of simulation instead of constant time as in RWM.

*2.2.5 Gauss Markov (GM) Mobility Model*

The need of the model, can make the node quicken, decelerate, or turn persistently. Liang and Haas (1999) have proposed this model and was generally utilized by Camp et al. (2002); Hu and Johnson (2000). This model expresses that the present movement of a node is identified with the previous movement through Gaussian conditions by utilizing speed, course, and Gaussian arbitrary noise. The level of reliance on previous speed and course is controlled by parameter α. This model is also known as feature temporal reliance. This model conducted towards the end of the simulation region can be expressed in various ways. Bai and Helmy (2004) have proposed a direction change towards the end without deciding decisively how this is done.

Amoussou et al. (2005) have proposed a 180° turn, this technique is difficult since it conflicts with one of the primary reasons behind why the GM model was proposed in any case; which was to maintain a strategic distance from sharp turns. Likewise, the transient reliance of GM is lost here, since the following course has nothing to do with the past one for this situation. Alenazi et al. (2013) have proposed an enhanced 3D GM model, which incorporates a buffer zone to allow MNs to move easily around the simulation boundaries without bounces and it works legitimately on a two-dimensional simulation region.

*2.2.6 Probabilistic Random Walk Mobility Model*

In this model, (Roy, 2010; Camp et al., 2002) Probability assumes a major role in the nodes positioning. A node has several movements like forward, invert, or remain in x and y-direction, which depends on the probability characterized in probability framework. PRW Model has three conditions: 0 (current position), 1 (Past position), and 2 (next position). In the probability framework, p(a, b) characterizes the node development from state a to b. The values which are defined will deny movements between the past and next positions of the MN without going through the present area.

This model is really more than purely random movements, but choosing the appropriate values of P(a, b) may prove difficult. The MN moves in straight lines for different periods of time and does not show the highly variable direction seen in the RWF Mobility Model.

*2.2.7 Boundless Simulation Area (BSA) Mobility Model*

This mobility model is different from other mobility models. A connection exists between the previous condition and the present condition of the node (Hass, 1997). In this model, the new direction and speed values are subject to the past direction and speed estimations of the node. Every time new values of the direction and speed are given.

At the point when a node reaches the end of the simulation area, it will not reflect again into the simulation region; rather, it keeps on traveling and returns on the opposite side of the re-enactment zone. This mobility model makes more practical node development since it relies upon the past speed and direction of the node. In the BSA model, Sharp turns and sudden stops can be disposed.

*2.2.8 City Section Mobility Model*

CSM model (Davies, 2000) is one of the important mobility models, which sets parameters such as the speed limit of every node way. Every MN starts the simulation at the limited intersection of two node ways. By going to the destination, it includes one flat and one vertical movement.

When the destination arrived, MN stops for some random time and after that picks another destination randomly and the procedure is repeated. This procedure is known as an epoch. The CSM model states that every MN following predefined ways will raise the average hop count in the simulation field. This model gives sensible movements compared to other mobility models.

*2.3 IoT Routing Protocols Requirements*

Home Automation, Urban-Low Power, and Lossy Networks (LLNs) are distinctive applications that have diverse qualities regarding movement, network size, and level of mobility. Subsequently, the necessities of the routing protocol contrasts starting with one application then onto the next application. In spite of these distinctions, the advantages are grouped into six categories (Martocci et al., 2010).

- *Mobility:* Generally IoT does not encounter movements, but rather a few applications require the movement of the nodes to be considered. A reasonable routing protocol ought to have the capacity to adapt to the area changes of the nodes.

- *Power Efficiency:* The nodes in the IoT environment are battery-driven, which run self-sufficiently. A routing protocol that is ingenious as far as power utilization is crucial to the usefulness of an IoT-based system.

- *Traffic Patterns:* A routing protocol for the IoT needs to coordinate the movement of its zone of sending.

- *Scalability:* The measure of the network in IoT may shift from 100 to 1000000 nodes. Consequently, the routing protocol intended for the IoT should coordinate the necessities of the system size.

- *Directionality:* In wireless networks, bi-directional connection between joins are not ensured. A routing protocol for the IoT must have the capacity to perceive and keep away from unidirectional connections at any rate and might have the capacity to utilize them one way.

- *Transmitter usage:* Concerning power utilization, the transmitter is the most costly part of a device. Routing Protocols may be outfitted with a number of components influencing the way in which they settle on routing.

## 3. Methodology

*3.1 Simulation*

In this paper, the simulations were done in Cooja simulator 2.7. Comparing to past versions, this Cooja version is more stable, which has been widely assessed in various works, (Dunkels et al., 2011; Eriksson et al., 2009; Kugler et al., 2013). The generated results are the average of 20 simulations on random topologies with various positions of the sink nodes and sender nodes, which are picked randomly in the simulation region. UDGRM is chosen in the physical layer. To simulate the scenario, RPL-collect is implemented. The sink node establishes a connection with the sender nodes in three phases:

Phase 1: Initializing the RPL DAG

Phase 2: Setting up a UDP connection

Phase 3: Printing received packets from the sender on stdout

For sender nodes, the sink sets up UDP connection and then starts to send a packet to the sink periodically. At the point when the sink node begins the instating, it tells the other nodes "I am the sink" and after that sends a DIO message intermittently. After each transmission, a time gap is expanded with the assistance of a trickle timer.

*3.2 Objectives of the Simulation Study*

In this paper, the simulations are varied in two types of networks: small scale and large scale networks. In the simulation region, 50 SKY motes were placed randomly. Due to PC capacity and SKY mote constrained resources, less number of motes were used. Transmission range of the motes is set to 100 m, and the Unit Disk Graph Medium (UDGM) model is utilized as a part of the simulations. In the network topology, sink motes were used for receiving packets from sender motes. the multipoint-to-multipoint method is used in the topology. The main objectives of these simulations are:

- To investigate the network behavior of RPL and utilizing RPL in static networks for power consumption while keeping its quality of service (QoS) in the network topology.

- To investigate the network behavior of RPL in a mobile

environment. In which we are evaluating the RWP mobility model and to calculate its power consumption.

- Comparison of static model with the RWP model using RPL routing protocol for investigating of the impact of these models and their power consumptions.
- To prove that subject indeed impacts the performance of RPL in mobile LLNs.

### 3.3 Simulator

In this area, Cooja simulator which is part of Contiki OS toolset qualities are outlined and a review of Cooja simulator 2.7, in view of SKY Mote is described.

### 3.4 Cooja Simulator

Cooja is a network simulator which is part of Contiki operating system toolset. Cooja is an open source and compatible with the need for analysis. Cooja's main feature is that it can simulate each node separately using hardware (Sky mote, z1 mote, etc) or software. Cooja can operate in three levels: Network Level, Operating System Level, and Instruction Level. In Cooja, extensions can also be added. Another main feature of Cooja is that a binary image of a platform can be evaluated in Cooja simulator like a virtual node.

By default, Cooja cannot support any mobility models. In this paper, Bonnmotion (Aschenbruck et al., 2013) is added to Cooja to generate the mobility patterns. The generated mobility patterns were converted into required data for further analysis which are obtained by Bonnmotion to a different format (WML).

### 3.5 Metrics of Interest

There are a couple of metrics that influence the routing mechanism in networks. Some of the particulars in RPL make it difficult in its utilization concerning some of the devices. There are five fundamental metrics used for RPL assessment are listed below:

- *Control Traffic Overhead:* In this metric, the control messages which are transmitted by nodes to gather DODAG and best parent between neighbors will be chosen.

- *ETX (Expected Transmission Count):* This metric represents the extreme number of re-transmissions of a separate packet to be effectively conveyed to the destination over a wireless connection.
- *Hop Count:* in this metric, the hop count represents the count of hops between nodes and root nodes.
- *Packet Delivery Ratio:* PDR is used to calculate the successful transmission of the packets from source to destination. Higher PDR indicates the better performance.
- *Node Power:* This metric is used to calculate the average power consumption of the nodes in the network. The formula for calculating the power of nodes (Lamaazi et al., 2016) is :

Power (mJ) = (Transmit/19.5 mA + Listen/21.5 mA + CPU_time/1.8 mA + LPM/0.0545 mA)/3V/(32768)

where,

CPU- represents the power consumption of the nodes which are in full power mode;

LPM- Listen Power Mode (LPM) represents the power consumption in low power mode;

Transmit- represents the transmission operations;

Listen- represents the listening operations;

Overall Power- represents the overall power consumption of nodes in the network.

This paper mainly concentrates on the node power mode as IoT mainly deals with resource-constrained environment and node power is used for evaluation of the RPL.

### 3.6 Simulated Scenarios

The test run configurations used in this paper are displayed in this section. Evaluation of RPL performances under the static and RWP models in only multipoint-to-multipoint topology was observed. To display the impact of RPL in different scenarios, the densities of the network are varied from 20 to 50 nodes on a scale of 10 (i.e., 20, 30, 40, and 50). In this paper, two types of nodes are utilized: sink nodes and sender nodes. The sink node is used to receive that data from sender nodes. The simulation scenario for Static Network is depicted in Figure 3.

Figure 4 depicts the simulation environment with RWP mobility model. In stand out from the dominant part of past examinations that attempt to study about RPL under mobility, the RPL performances are assessed under static and RWP mobility model to demonstrate how RPL works when the mobility is presented and which ones give better performance results as far as power consumption is concerned.

4. Results

Power consumption is the principal constraint of WSN, and it is crucial to assess the power consumption of a WSN running RPL. The assessment work was conducted in two perspectives. Initially, the power consumption of the entire system is assessed to get a general picture of power. At that point, the power consumption of individual nodes is assessed to get more bits of knowledge. The distinction between power consumption between client nodes and sink nodes is additionally assessed and analyzed.

In WSN nodes, radio transceiver is the main reason for power consumption. For example, power consumption by the radio is three solicitations greater than CPU performance of Z1 node. Therefore, the main focus will be on radio transmission to calculate the power consumption. Contiki supports many MAC protocols.

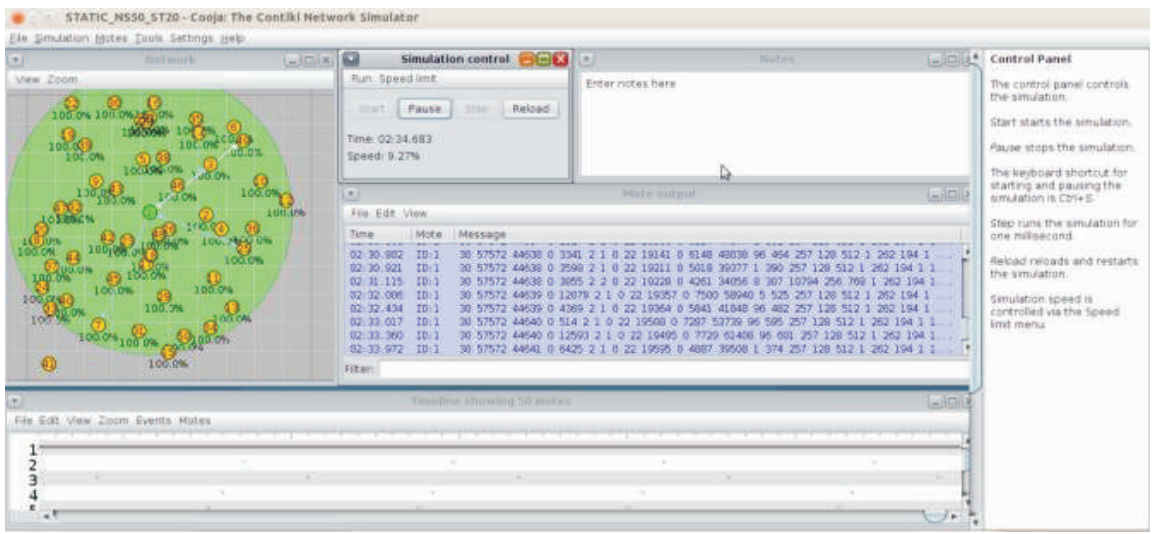

Figure 3. Test Run Simulation of Nodes in Static Model

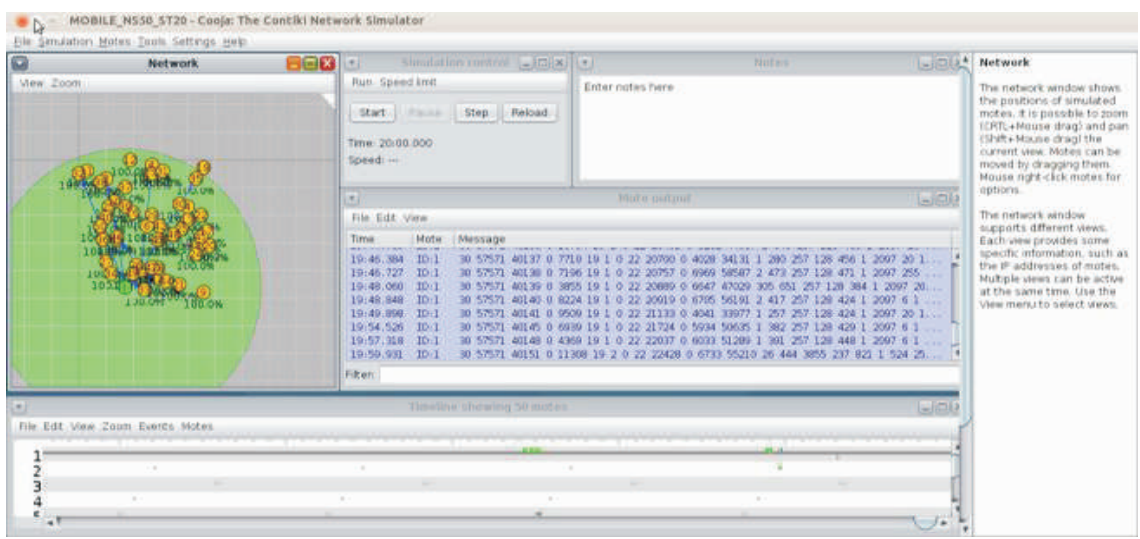

Figure 4. Test Run Simulation of Nodes in the RWP Model

One of the important protocols is the Contiki MAC, which is also the default Radio Duty Cycle (RDC) protocol in the Contiki operating system. RDC protocol is used to make the nodes to be standby mode for power saving. Hence, radio is utilized as a performance metric for power level comparison and the readings will be calculated in joules. Most of the time radio will be in a working mode so an efficient power consumption protocol is required.

From the observations, sink nodes spend most of their power on receiving packets from sender nodes without efficiency. More power consumes in sender nodes than sink nodes for sending packets rather than power consumption in receiving packets. Therefore, RPL control messages were examined and suggests that if packets direction work in stable mode power consumption may reduce if the DAO messages in RPL control messages were cut down.

*4.1 Power Consumption*

In this paper, different network densities are varied to measure the consumed power in different power scenario. Figure 5 shows that the CPU power of the RWP network consumes more power than a static network when it becomes denser. This increase is mainly due to the number of transmissions of packets sent by nodes. The augmentation of the number of nodes from 20 to 50 nodes provides an augmentation of the power of 19%. Indeed, an unsuccessful transmission provides an augmentation of power consumption as opposed to a successful one.

Low Power Mode (LPM) is one of the important features of the WSNs. LPM is used to reduce the power consumption of the sensor by going to standby mode. Figure 6 demonstrates the correlation of the power utilization in nodes while in LPM mode. In WSNs, Low-power listening is a duty cycling mechanism where the receivers occasionally turn on their radios to survey the radio medium for action. In the event that there is action, the radio is kept on for a more extended time on the off chance that a packet would be sent to the node.

Before sending a packet, the sender sends various strobe packets with the expectation to tell the receiver that a neighbor needs to send a packet. Since all neighbors are intermittently inspecting the radio medium, they will see the strobe packets and keep their radios on fully expecting the data packet.

Figure 7 shows the comparison of Nodes Listen Power in the two models. The strobe time frame is characterized to be the length of the sleep period of the neighbors so the neighbors will be woken up by the strobe packets. There are a few variations of low-power listening (Herberg & Clausen, 2011). Some don't send strobe packets to awaken their neighbors yet send a long wake-up tone. This tone fills an indistinguishable need from the strobe packets yet cannot contain any data (Radio duty Cycling, 2015).

Low-Power Transmit is a method that turns around the parts from low-power listening. Rather than having the receivers occasionally test the radio medium, the receivers intermittently transmit a test packet into the radio medium (Broch et al., 1998). To send a packet, the sender turns on its radio and tunes in for a test packet (probe packet) from the neighbor that ought to get the data packet. At the point when the test packet arrives, the sender instantly sends its data packet. The receiver node, which keeps its radio on for a brief span in the wake of sending its test packet, will along these lines get the data packet.

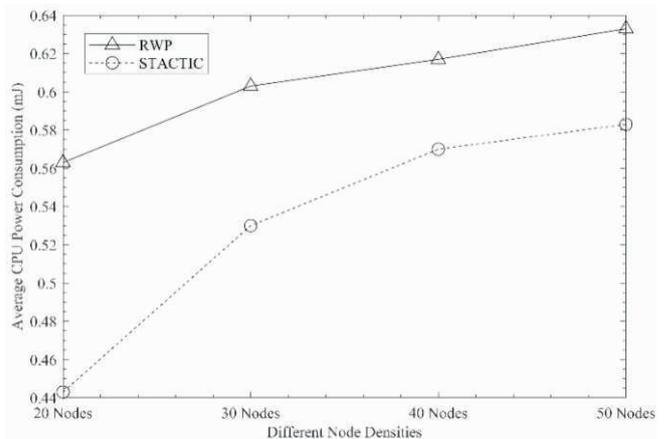

Figure 5. Comparison of CPU Power Consumption for Different Densities of Network

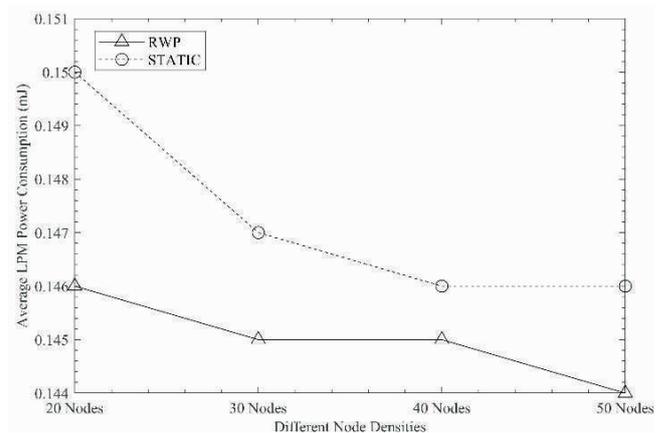

Figure 6. Comparison of LPM Power Consumption for Different Densities of Network

This technique executes just at the network's nodes. The wake-up activity is started by a passage which empowers its radio acknowledgment and begins tuning in for transmissions from network nodes. Figure 8 illustrates the comparison of Low-power transmit mode for different network densities (Royer et al., 2001).

As explained above, the introduction of mobility to the nodes allows consuming fewer resources. The authors have measured the average power consumption in such case to prove this conclusion as illustrated in Figure 9.

The introduction of mobility into the nodes increases the power consumption by 36.08%. Sink nodes do not consume much power because they work as collectors of data from senders and do not need to retransmit packets to check the network availability or to find neighbors. The sink nodes transmit the first message to all nodes to be presented as a sink. After this message, sink nodes act as a server such that all sender nodes try to send to them their collected data. These activities legitimize the way that the power consumption is increased if there should be an occurrence of mobility when contrasted with the static situation (Lamaazi et al., 2017).

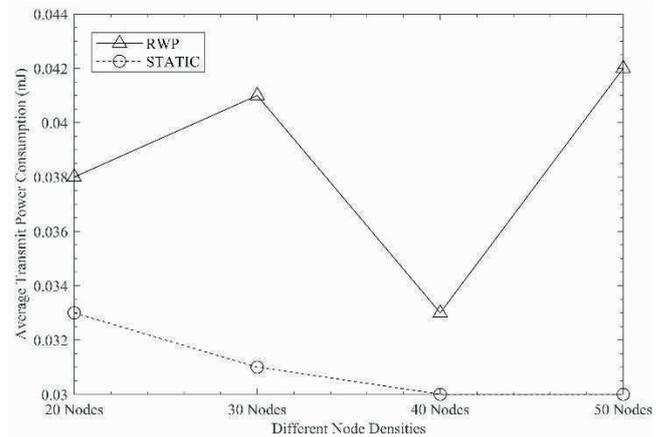

Figure 8. Comparison of Low-Power Transmit Mode for Different Densities of Network

At the point when the sink node communicates the DIO messages, sender nodes in a similar transmission scope of the sink get the DIO messages. A short time later, nodes in light of the OF choose to join the DODAG. The sink node broadcasts the DIO messages every now and then when the network is not steady or on account of another node joins the network system.

Hence, the introduction of mobility model makes the network unstable as the sink and other nodes move in a random manner making them move closer or farther. This makes the DODAGs reformation, increase in the number of retransmission of expected transmission count (ETX) and exchange of control messages (DIO and DAO) so frequent. The above discussion makes it clear that there is an increase in control messages which obviously increases power consumption.

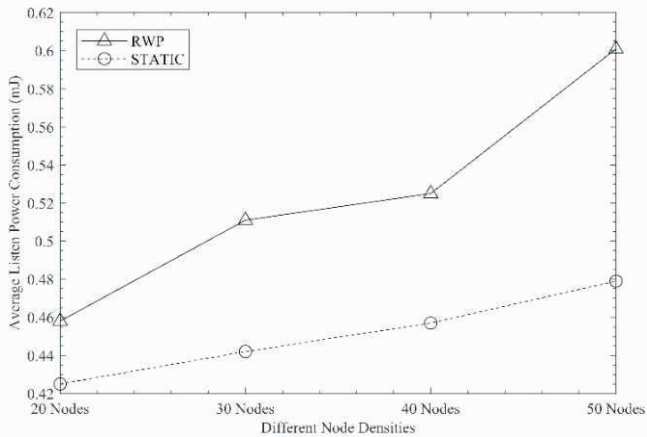

Figure 7. Comparison of Low-Power Listen Mode Power Consumption for Different Densities of Network

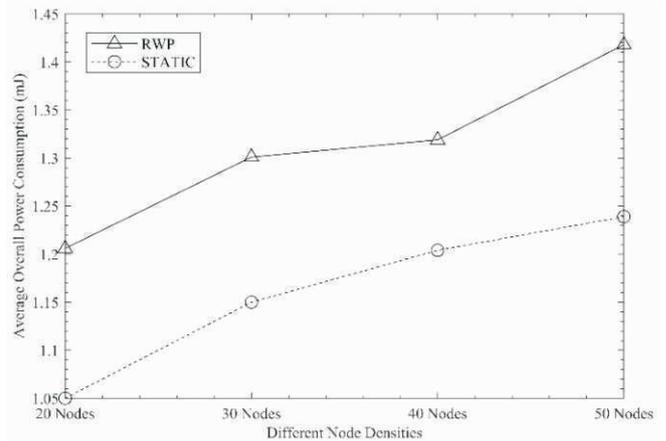

Figure 9. Comparison of Overall Power Consumption for Different Densities of Network

In the present work, two imperative parameters to consider the conduct of RPL: the node count (Network Size) and the Mobility Models. In all figures, the simulation analyzes the performance of RPL, considering only one output parameter namely Power Consumption. The topology used in all simulations is "one-to-many", which contains one sink node, with the rest being sender nodes.

Conclusion

The performance of any wireless routing protocols relies upon the period of interconnections between any two nodes that are exchanging data and also on the time duration of interconnections between nodes of the path containing n-nodes.

The mobility of the nodes influences the number of connected paths, which thus influence the execution of the routing protocol/algorithm. In this paper, the effect of network scalability and mobility of nodes on the power consumption of IoT routing protocol RPL is discussed and concludes that mobility of nodes influences the performance of the routing protocols.

Future Scope

However, in future, it is expected that various applications with diverse topography and node configuration are common in IoT environment. Broadly fluctuating mobility characteristics, for example, Group Mobility Models and Entity Mobility Models are relied upon to significantly affect the performance of the various routing protocols.

In future, the performance of the RPL protocol with respect to various mobility models can also be analyzed. A comparative study of various mobility models on the performance of RPL can be analyzed.

In addition, it is expected that some computing techniques can be used to reduce the power consumption with different models, different mobility speeds, and network densities.

## ABOUT THE AUTHORS

*Chandra Sekhar Sanaboina is currently working as an Assistant Professor in the Department of Computer Science and Engineering at Jawaharlal Nehru Tehnological University, Kakinada, Andhra Pradesh, India. He is presently pursuing his Ph.D. under the guidance of Prof. Pallamsetty Sanaboina from Andhra University, India. He obtained his B.Tech in Department of Electronics and Computer Science Engineering and M.Tech in Department of Computer Science and Engineering from Vellore Institute of Technology, India. He has over 10 years of teaching experience and his areas of interests include Wireless Sensor Networks, Internet of Things, Machine Learning, and Artificial Intelligence.*

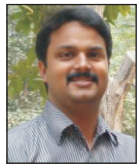

*Dr. Pallamsetty Sanaboina is currently working as a Professor in the Department of Computer Science and Systems Engineering at Andhra University, Visakhapatnam, Andhra Pradesh, India. He has completed his Ph.D. in the area of Wireless Sensor Networks from Andhra University, Visakhapatnam, India. He has 28 years of teaching and Research experience. He has over 200 publications in International Journals and Conferences of repute. More than 200 Master's students and 18 Doctoral students had completed their degree so far under his supervision, and currently supervising 8 Doctoral candidates in the area of Mobile Ad-hoc Networks, Wireless Sensor Networks, Ubiquitous Computing, and Web of Things which are his research interests.*

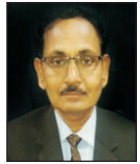